\documentstyle[emulateapj]{article}

\def\lae{\mathrel{<\kern-1.0em\lower0.9ex\hbox{$\sim$}}}
\def\gae{\mathrel{>\kern-1.0em\lower0.9ex\hbox{$\sim$}}}

\slugcomment{\smallskip Accepted to the Publications of the Astronomical Society of the Pacific}
 
\begin{document}
 
\title{The Nature of the Red Giant Branches in the Ursa Minor and Draco 
Dwarf Spheroidal Galaxies\altaffilmark{1}}
\altaffiltext{1}{This work is based on observations obtained with the Hobby-Eberly Telescope, 
which is a joint project of the University of Texas at Austin, the Pennsylvania
State University, Stanford University, Ludwig-Maximillians-Universit\"at 
M\"unchen, and Georg-August-Universit\"at G\"ottingen.}

\medskip

\author{Matthew D. Shetrone}
\affil{McDonald Observatory, University of Texas, P.O. Box 1337, Fort Davis, Texas 79734 \\
   {\rm shetrone@astro.as.utexas.edu}}

\smallskip

\author{Patrick C\^ot\'e}
\affil{Department of Physics and Astronomy, Rutgers University, New Brunswick, NJ 08854 \\
   {\rm pcote@physics.rutgers.edu}}

\and
\author{Peter B. Stetson}
\affil{Dominion Astrophysical Observatory, Herzberg Institute of Astrophysics, National 
Research Council, 5071 West Saanich Road, Victoria, BC, V9E 2E7, Canada \\
   {\rm Peter.Stetson@nrc.ca}}

\medskip

 
 
\begin{abstract}
Spectra for stars located redward of the fiducial red giant branches of the Ursa Minor 
and Draco dwarf spheroidal galaxies have been obtained with the Hobby-Eberly 
telescope and the Marcario Low Resolution Spectrometer.
From a comparison of our radial velocities with those reported in previous
medium-resolution studies, we find an average difference of
10 km s$^{-1}$ with a standard deviation of 11 km s$^{-1}$.
On the basis of these radial velocities, we confirm the membership
of five stars in Ursa Minor, and find two others to be nonmembers.
One of the confirmed members is a known carbon star which lies redward of RGB;
three others are previously unidentified carbon stars.
The fifth star is a red giant which was found previously by Shetrone et~al. (2001) 
to have [Fe/H] =$-1.68\pm0.11$dex.
In Draco, we find eight nonmembers, confirm the
membership of one known carbon star, and find two new
members. One of these stars is a carbon star, while the
other shows no evidence for C$_2$ bands or strong atomic bands, 
although the signal-to-noise ratio of the spectrum is low. Thus, we find no 
evidence for a population of stars more metal-rich than [Fe/H] $\simeq -1.45$ dex 
in either of these galaxies.  Indeed, our spectroscopic survey suggests that 
every candidate suspected of having a metallicity in excess of this value based 
on its position in the color-magnitude diagram is, in actuality, a carbon star. 
Based on the census of 13 known carbon stars in these two galaxies, we estimate 
of the carbon star specific frequency to be
$\epsilon_{\rm dSph} \simeq 2.4\times10^{-5}$ $L_{V,{\odot}}^{-1}$,
25-100 times higher than that of Galactic globular clusters. 
\end{abstract}
 
 
\keywords{galaxies: abundances --- galaxies: dwarf --- 
galaxies: individual (Draco, Ursa Minor) --- stars: carbon}
 
 
%

\section{Introduction}

To fully appreciate the Closed or Leaky Box nature of dwarf galaxies,
their full nucleosynthetic and star forming histories must be established. Doing so
requires that stars lying in the extreme low- and high-metallicity
tails of their host galaxies be found and studied. Unfortunately, the coupling of 
age and metallicity with position on the red giant branch (RGB) makes
a photometric search for such stars difficult. Even for systems characterized
by a single burst of star formation, identifying bona fide metal-rich stars
on the basis of photometry alone is challenging since carbon stars ---
by virtue of their strong molecular absorption in the ultraviolet and the resulting
flux redistribution to longer wavelength --- have 
colors which can mimic those of more metal-rich objects. The carbon
stars belonging to Pop II systems are usually of
the CH-type (Aaronson et~al. 1983), named for the spectral characteristics of metal-poor but
carbon-rich stars.  It is commonly believed that they result from mass
transfer in compact binary systems (McClure 1984). Thus, carbon stars
provide insights into binary formation and evolution.

The Ursa Minor and Draco dwarf spheroidal (dSph) galaxies have long been
suspected of showing an intrinsic dispersion in abundance (e.g., Zinn 1978). 
Recently, high-resolution spectroscopy for a dozen stars in these galaxies
has provided unmistakable evidence for an intrinsic spread in metallicity
in both objects:
i.e., $\Delta$[Fe/H] = 0.73 and 1.53 dex for Ursa Minor and Draco, respectively
(Shetrone, C\^ot\'e \& Sargent 2001).
In each case,
the most metal-rich member was found to have [Fe/H] $\simeq -1.45$ dex.
However, recent photometric studies of these galaxies reveal a number of objects
which fall redward of their fiducial giant branches, so it is conceivable
that their metallicity distribution functions extend to still higher
metallicities. This is particularly true of Ursa Minor, which has been
the subject a proper motion survey by Cudworth, Seitzer \& Majewski
(2002). These authors find a small number of ostensible members whose
location in the color-magnitude diagram (CMD) makes them prime candidates
for metal-rich stars.

In this paper, we present the results of a small survey of stars redward of 
the fiducial giant branches of the Ursa Minor and Draco galaxies to
determine whether these objects are metal-rich RGB stars, carbon stars or 
nonmembers. We find no evidence for a population of stars more metal-rich
than [Fe/H] $\simeq -1.45$ dex in these galaxies. Every radial velocity
member which falls further to the red of the RGB fiducial sequences,
and for which adequate spectrosopic material is available, is found to
be a carbon star.

\section{Observations and Reductions}

Program stars in the direction of Ursa Minor and Draco were selected
from the photometric catalogs of Stetson (2002)\footnote{The Draco 
photometry in this
paper is based on analysis of archival CCD material during the course of
preparing the paper Stetson (2000).  The data used here
are on the photometric system of that paper, and more details about the
Draco and other fields can be obtained from the Web site
http://cadcwww.hia.nrc.ca/cadcbin/wdb/astrocat/stetson/query/} and
Cudworth et~al. (2002), respectively. Since this latter study
also reports proper motion measurements, we only selected 
stars with membership probablities in excess of 50\%. Since our aim was
to establish the full range in metallicity spanned by stars in these galaxies,
we targeted objects displaced from the canonical red giant branches (RGBs),
with particular emphasis on the red, potentially metal-rich objects. Note 
that some stars redward of the fiducial RGBs have published radial velocities
(e.g. Armandroff, Olszewski \& Pryor 1995; hereafter A95), and a few are classified as carbon
stars (Aaronson et~al. 1982; Azzopardi et~al. 1986; Olszewski et al. 1995;
Aaronson et~al. 1983; Canterna \& Schommer 1978; Zinn 1981, A95).
We re-observed a number of these stars to
confirm the precision of our radial velocities, to use as carbon star 
abundance standards, and to look for weak carbon star features missed by 
previous, lower signal-to-noise analyses.

The observations were conducted at the Hobby-Eberly telescope during the second 
queue observing period of 2000. The Marcario Low Resolution Spectrograph (LRS; Hill 
et ~al. 1998) was used in its highest resolution mode --- a 600 l/mm grating, 
1$^{{\prime\prime}}$ slit, and 2x2 binning --- to produce a resolution of $R$ = 
1200. In this configuration, a 0.1 binned pixel
velocity precision would be 25 km s$^{-1}$.  The program was conducted
in a wide range of seeing conditions, lunar phases and atmospheric 
transparencies. Despite the variety of observing conditions,
the spectra had typical signal-to-noise ratios of S/N $\sim$ 25, 
obtained in single exposures of less than 20 minutes; the total time 
spent with the shutter open on the sky for this program was four hours.
For each spectrum, flat-field and line calibration lamps were 
taken. The spectra were reduced with standard 
IRAF\altaffilmark{2}\altaffiltext{2}{IRAF is distributed by the National Optical
Astronomy Observatories, which are operated by the Association of Universities for
Research in Astronomy, Inc., under contract to the National Science Foundation.} 
long-slit extraction packages.  The initial wavelength scales were applied 
using HgCdZn and Ne lamps.

\section{Radial Velocity and Abundance Analysis}

Velocity zero-point corrections were made by using the many weak night 
emission lines in this portion of the spectrum (Osterbrock et~al. 1996). 
The spectrum of each star was inspected visually to identify any strong 
molecular features such as TiO, CaH, C$_2$, CN, or CH. For those stars
lacking strong molecular features, a synthetic spectrum
was created using the colors given in Table 1 and our own color-temperature-gravity
relationship for an assumed metallicity of [Fe/H] = $-$1.8 dex.

A synthetic template spectrum was generated for those stars with visible molecular 
features. For objects with visible TiO and CaH features, we created a solar
metallicity synthetic spectrum, using an incomplete TiO line list.
Stars showing C$_2$, CN, and CH features, were modeled using a synthetic spectrum with
[Fe/H] = $-$1.8 dex and an enhanced carbon abundance: i.e., $\log\epsilon$(C) $>$ $\log\epsilon$(O).
Radial velocities were measured via cross-correlation with these templates.
Previous radial velocity analysis (e.g. A95 and Olszewski et~al. 1995)  
employed radial velocity standards instead of synthetic templates.  Analyses
which employ radial
velocity standards are able to remove some intrinsic systematic 
instrumental biases but sometimes suffer from flexure zero point errors and 
spectral type mismatches.  Using synthetic templates and the night time 
emission or absorption lines 
allows one to side step the latter problems but does not give any information
about the former.   
A similar technique was employed in Shetrone (1994) with high resolution
spectra.    The use of synthetic templates for cross-correlation is also
useful for programs which are queue scheduled on large telescopes when 
instrument setups change hour to hour and the data is acquired over the 
course of months and telescope time does not need to be spent on bright 
radial velocity standards.
Data for the confirmed members of Ursa Minor and Draco are reported in Table 1, which 
gives the star ID, previous identifications following the naming scheme in A95, right 
ascension, declination, S/N per 2 pixel resolution element, and LRS radial 
velocity. Comments concerning the classification of the stars are given in 
the final column.

The upper panel of Figure 1 compares our LRS velocities with those measured 
by A95. There is a systematic offset of 10 km s$^{-1}$ with a standard deviation of 
11 km s$^{-1}$ around that offset. The lower panel of Figure 1 shows the velocity
differences, ${\Delta}v_r = v_r({\rm LRS}) - v_r({\rm A95})$, plotted against S/N 
per resolution element. If the offset in these differences is taken to be the
result of zero-point difference, then we find that LRS in its highest resolution
mode can achieve a velocity accuracy of $\sim$ 15 km s$^{-1}$. Recall that a 0.1 binned pixel
velocity error would be 25 km s$^{-1}$, as indicated by the dashed lines.

A comparison was made between the synthetic carbon star spectrum and those
of known carbon stars in Ursa Minor and Draco. Although the absolute
carbon and nitrogen abundances from this {\it ad hoc} model can not be trusted,
the carbon abundance is higher than the oxygen abundance, 
and the fit is visually similar to the observed spectra of the known
carbon stars. For each star, we created a  pair of synthetic spectra: i.e.,
one with, and one without, enhanced carbon abundances. For the
coolest stars, the C$_2$ features are the strongest features in the
spectrum; among the hotter stars, the C$_2$ features are similar in
strength to the atomic features. Given the signal-to-noise
of our spectra, we are able to detect the hot carbon
stars that previous investigations would have missed. 

Figure 2 shows our LRS spectra for five stars belonging to Ursa Minor. The sample 
includes one previously recognized carbon star, three new carbon stars, and one 
apparently normal RGB star.  In this galaxy, we therefore confirm the carbon star 
identification of one giant, UM152 (30614), and detect several additional 
carbon stars: UM1545 (37759, N42), UM1859 (35869), and UM536 (32961, JI12).  
In Draco, we confirm the classification
of one carbon star, J (C2, 20733), and identify one new carbon star: 68.
A complete list of the 13 carbon stars in these two galaxies is
presented in Table 2.

The radial velocity members of Draco which fall on the galaxy's fiducial giant 
branch 
are 195, S37 and M.\altaffilmark{3}\altaffiltext{3}{For Draco, the IDs are from
Armandroff et al. (1995) and Baade \& Swope (1961) unless preceeded by an ``S" in which 
case they refer to the catalog of Stetson (2002).  For Ursa Minor, the IDs are from
Cudworth et~al. (2002). When available, cross-identifications for these stars 
are given.} All stars with visible TiO and CaH bands were found to be 
nonmembers.  We find eight nonmembers which fall redward of the RGB in Draco,
including two stars (90 and 193) previously identified as nonmembers from their proper 
motions (Stetson 1980). A single star located to the blue of the RGB was also found to 
be a nonmember.  Likewise, two stars located redward of the RGB in Ursa Minor were 
found to be nonmembers.

For Ursa Minor, only one radial velocity member in our sample
has an abundance pattern which is definitely not that of a carbon star. This
star, UM1846 (347, 35606, 297Q3), was studied by Shetrone et~al. (2001) as \#297 
using high-resolution 
spectra from the Keck I Telescope. They found it to have [Fe/H] = $-$1.68$\pm$0.11 
dex. Using our low resolution spectra we can constrain the abundance to 
[m/H] = $-1.8 \pm 0.3$. With the exception of this star, we note that all of the
proper motion members located redward of the RGB in this galaxy are now classified
as carbon stars. This situation may also hold in Draco although there are a number
of additional objects located redward of the RGB for which spectroscopy is not
available. Unfortunately, our spectrum for star 194 in Draco lacks the S/N needed to determine 
if it too is a carbon star. We can merely constrain the abundance of this star to be
[m/H] $< -1.1$ dex based upon the depth of the blended atomic features. 
Additional spectroscopy for this star, as well as for the remaining
objects located on the red side of the Draco RGB, might prove profitable.

In Figure 3, we show CMDs for both Ursa Minor and Draco. In the right panel,
radial velocities of Draco from A95 and our LRS survey have been used to separate members
from nonmembers in the photometric catalog of Stetson (2002). The CMD of
radial velocity and/or proper motion members in Ursa Minor is shown in the
left panel of this figure. The new carbon stars are indicated by the 
filled stars, while the previously known carbon stars are denoted by the filled squares. 
In both galaxies, the new objects fall below the tip of the RGBs ---
similar to the known carbon stars which, from infrared
photometry, have bolometric magnitudes of $M_{\rm bol} \gae -3.5$.
Thus, the bulk of these stars are almost certainly CH-type carbon stars (McClure 1985) 
rather than the more luminous N- and R-type carbon stars found in some Local Group 
dwarf galaxies which contain intermediate-age populations (Aaronson \& Mould 1985). 
Possible exceptions to this claim may be stars J (C2, 20733) and 461 in Draco, 
which 
we find to be photometric variables (at the 10 and 4$\sigma$ levels, 
respectively).
Such variability is common among asympototic giant branch (AGB) stars belonging
to intermediate-age populations. The existence of a small, intermediate-age component
in Draco has been suggested by Carney \& Seitzer (1986) and Grillmair et~al. (1998),
who noted the presence of a population of stars above the main-sequence turnoff. 
On the other hand, radial velocity monitoring of stars J (C2, 20733) and 461 
(Olszewski et~al. 1996) 
show no evidence for the variations which might be expected in {\sl either} scenario for
their enhanced carbon abundances: i.e., mass-transfer from evolved close secondaries
or dredge-up during AGB evolution, an evolutionary phase characterized by thermal
pulsations.

The solid curves in each panel of Figure 3 show 12 Gyr 
isochrones from Bergbusch \& VandenBerg (1992) having 
[Fe/H] $= -2.26~-2.03~-1.78~{\rm and}~-1.48~{\rm dex}$, and shifted according to the reddenings and distances
given in Mateo (1998).\altaffilmark{4}\altaffiltext{4}{In Draco, the most metal-poor star
found by Shetrone et~al. (2001) has [Fe/H] = $-2.97\pm0.15$ dex --- considerably lower
than the most metal-poor metallicity isochrone of Bergbusch \& 
VandenBerg (1992). However, the offset relative to the most-poor isochrone is expected 
to be relatively modest since, at low abundances, isochrone shapes are largely 
independent of metallicity.}
As mentioned above, almost every star which is radial 
velocity and/or proper motion member of these galaxies, and which
falls redward of the RGB expected for an old, [Fe/H] $\simeq -1.45$ population, 
is classified as a carbon star.

The three carbon stars discovered by Armandroff et~al. (1995) in these two galaxies
increased the number of such objects from six to nine. Our LRS survey brings their
census to $N^{\rm C}_{\rm dSph} = 13$. It is 
interesting to compare the frequency of carbon stars in these dSph
galaxies with those in Galactic globular clusters. C\^ot\'e et al. (1997)
identified a total of three stars in globular clusters which show strong
C$_2$ bands (two in $\omega$ Cen and one in M14), among a sample of $N_{\rm GC} = 147$ 
clusters (Harris 1996). If the selection criterion is relaxed to include hotter
stars which show strong CN and CH absorption but no C$_2$ bands, then 
$N^{\rm C}_{\rm GC} \sim 10$ (McClure 1984).
Since the mean luminosity of Galactic globular clusters is $\langle L_V \rangle \simeq 7.5\times10^4$
$L_{V,{\odot}}$, the specific frequency of CH stars in globular clusters, 
$$\epsilon_{\rm GC} = N^{\rm C}_{\rm GC}[N_{\rm GC}\langle L_V \rangle]^{-1},~\eqno{(1)}$$
is found to be
$$1\times10^{-6}~L_{V,{\odot}}^{-1}~\lae \epsilon_{\rm GC}~\lae 2.7\times10^{-7}~L_{V,{\odot}}^{-1}.$$
Mateo (1998) reports luminosities of 2.9$\times10^5$ and 2.6$\times10^5$
$L_{V,{\odot}}$ for Ursa Minor and Draco, respectively. Thus, the
specific frequency of CH stars in these galaxies is 
$$\epsilon_{\rm dSph} \simeq 2.4\times10^{-5}~L_{V,{\odot}}^{-1},$$
or 25-100 times higher than that for globular clusters. 
Since field CH stars are known to be the end-products of mass
transfer between contact binaries (McClure 1984), the disparity between
these specific frequencies may be evidence that in the denser
globular-cluster environments, binaries are either disrupted or driven
toward orbital shrinkage and eventual coalescence long before the primary
can become an AGB star and experience carbon dredge-up and subsequent mass
transfer.  Binary systems in the looser dwarf spheroidal systems and in the
field, on the other hand, can normally survive long enough to reach this
phase of evolution.
Renewed searches
for additional CH stars in Local Group dwarf galaxies might prove 
worthwhile, given that the census of such stars in Ursa Minor and Draco,
once thought to be complete, has increased 
by nearly 50\%.

\section{Summary}

We have used the Hobby-Eberly telescope and Marcario LRS to carry out a search for metal-rich stars 
in the Ursa Minor and Draco dSph galaxies. From a comparison of our radial
velocities for stars in common with those targeted by previous higher-resolution studies, we
find that LRS is sufficiently stable to achieve a radial velocity accuracy of approximately 1/15 pixel, 
or 15 km s$^{-1}$ at S/N $\simeq$ 25 per resolution element.
Although a number of members were found redward of the fiducial
RGB in this survey, there is no evidence for a population more metal-rich
than [Fe/H] $\simeq -1.45$ dex in these galaxies.  Indeed, nearly every radial velocity 
member redward of the fiducial RGB is found to be a carbon star.  Combining our sample of 
four new carbon stars with those previously known, we find a total of seven
carbon stars Ursa Minor and six in Draco. Thus, relative to Galactic globular
clusters, these galaxies appear to be are overabundant in carbon stars by 
a factor of 25-100.

\acknowledgments
 
We extend our thanks to the staff of the Hobby-Eberly telescope and, in particular, 
to Grant Hill, Brian Roman, Gabrelle Saurage, and Teddy George for providing the high 
quality service observations that made this project possible.  Thanks also to 
Kyle Cudworth for providing the photometry for Ursa Minor prior to publication.

\clearpage
\plotone{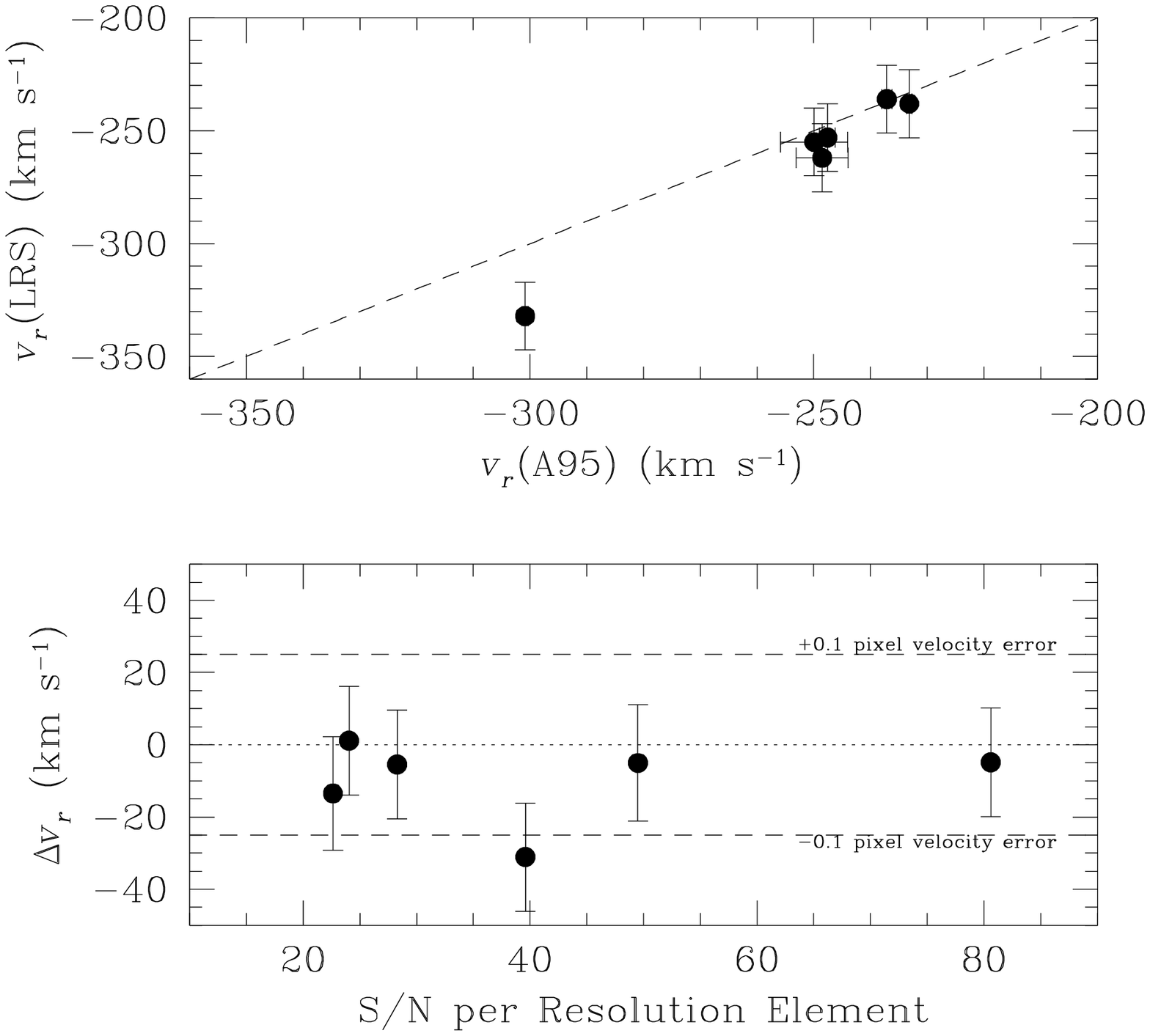}
\figcaption[shetrone.fig01.eps]{(Upper Panel)  Comparison of our LRS radial velocities
with those of Armandroff et~al. (1995; A95). The dashed line indicates the one-to-one
relation. The velocity uncertainty for our LRS measurements is taken to be
${\sigma}(v_r) = 15$ km s$^{-1}$ based on the results below. (Lower Panel) Radial
velocity difference, ${\Delta}v_r = v_r({\rm LRS}) - v_r({\rm A95})$, for stars
in common with Armandroff et~al. (1995).  The dashed lines show 0.1 binned pixel
velocity errors for LRS in this configuration; the distribution of points is
consistent with a 0.06 $\sim$ 1/15 binned pixel velocity error, corresponding to 15 km s$^{-1}$.
Error bars show this uncertainty added in quadrature to those reported by Armandroff
et~al. (1995).
\label{fig1}}
\clearpage

\plotone{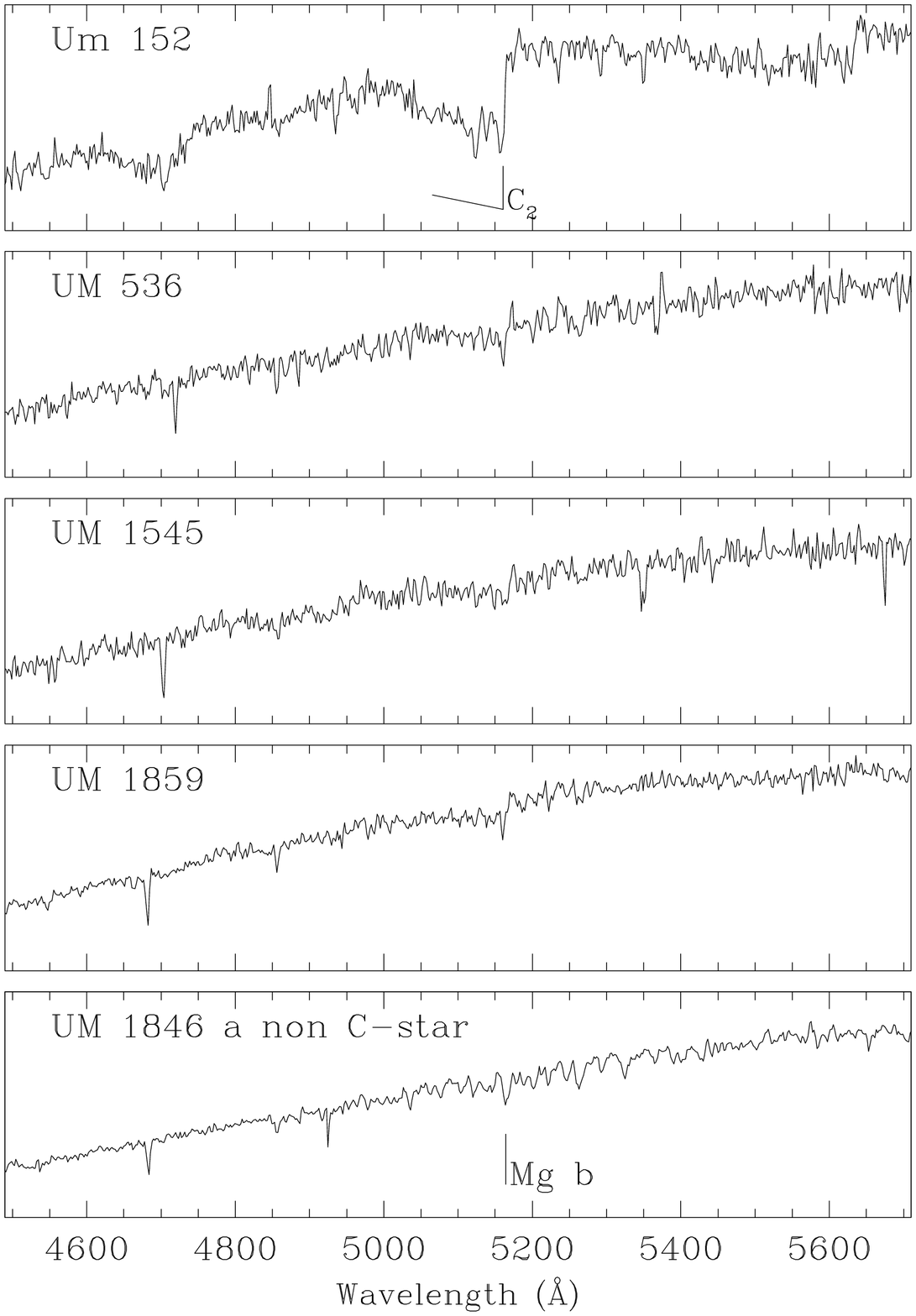}
\figcaption[shetrone.fig02.eps]{LRS spectra for radial velocity members of the Ursa Minor
dSph galaxy. UM 152 (30614) is a previously recognized carbon star; 
UM 536 (32961, JI12), 1545 (37759, N42)
and 1859 (35869) are  newly discovered carbon stars. The lower panel
shows the LRS spectrum of UM 1846 (347, 35606, 297Q3), a normal RGB star. 
The strong C$_2$ band
in UM 152 (30614) which was used to identify the C-stars is marked in the 
upper panel.
The Mg$b$ feature, which blend with the red most part of the C$_2$ band in 
C-stars, is marked in the lower panel.
\label{fig2}}
\clearpage

\plotone{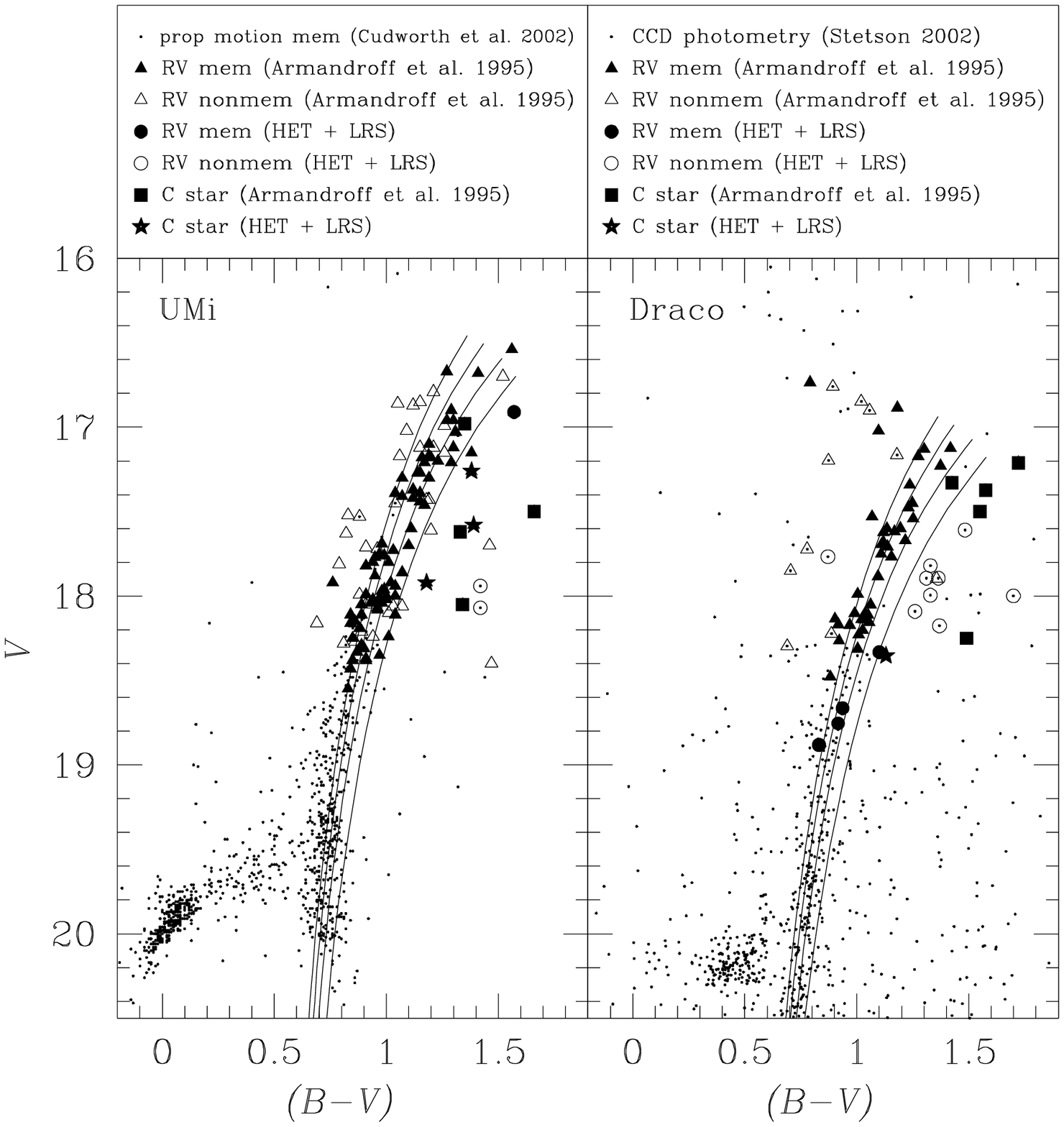}
\figcaption[shetrone.fig03.eps]{$BV$ color magnitude diagrams for the Ursa Minor and Draco
dSph galaxies, based on photometry from Cudworth et~al. (2002) and Stetson (2002),
respectively. The various symbols are explained in the legends at the top
of the panels. The solid curves in each panel show 12 Gyr isochrone from
Bergbusch \& VandenBerg (1992) having 
[Fe/H] $= -2.26~-2.03~-1.78~{\rm and}~-1.48~{\rm dex}$, 
and shifted according to the reddenings and distances given in Mateo (1998).
For comparison, Shetrone et~al. (2001) find weighted mean metallicities for
Ursa Minor and Draco of [Fe/H] $= -1.90\pm0.11$ and $-2.00\pm0.21$ dex, respectively. 
The most metal-poor star found by Shetrone et~al. (2001) in Ursa Minor has [Fe/H] $= -2.18$ dex, 
while the most metal-poor star in Draco has [Fe/H] $= -2.97$ dex. 
For both galaxies, the most metal-rich star in the study of Shetrone et~al. (2001)
has [Fe/H] $\simeq -1.45$ dex.
\label{fig3}}
\clearpage

\begin{deluxetable}{llccccl}
\tablecolumns{7}
\tablenum{1}
\tablewidth{0pc}
\tablecaption{Ursa Minor and Draco Member Stars Observed With LRS}
\tablehead{
\colhead{Name} &
\colhead{Other} &
\colhead{${\alpha}({\rm J}2000)$} &
\colhead{${\delta}({\rm J}2000)$} &
\colhead{S/N} &
\colhead{$v_r$} &
\colhead{Notes} \\
\colhead{} &
\colhead{} &
\colhead{} &
\colhead{} &
\colhead{} &
\colhead{(km s$^{-1}$)} &
\colhead{} 
}
\startdata
\multicolumn{7}{c}{Ursa Minor\tablenotemark{1}}\\
UM1859 & 35869             &15:08:08.99 & 67:09:21.4 & 49 & -255  & New carbon star    \\
UM1545 & 37759, N42        &15:08:35.67 & 67:03:41.1 & 24 & -236  & New carbon star    \\
UM536  & 32961, JI12       &15:11:36.35 & 67:18:07.2 & 28 & -253  & New carbon star    \\
UM152  & 30614             &15:11:55.47 & 67:25:08.6 & 23 & -262  & Known carbon star  \\
UM1846 & 347, 35606, 297Q3 &15:08:27.16 & 67:10:07.3 & 81 & -238  & RGB star           \\
\multicolumn{7}{c}{Draco}\\
J        & Draco C2, 20733 &17:20:00.68 & 57:53:46.7 & 40 & -332  & Known carbon star  \\
194      &                 &17:20:27.24 & 57:56:12.1 & 20 & -293  & RGB star           \\
68       &                 &17:19:57.29 & 57:55:04.7 & 25 & -246  & New carbon star    \\
195      &                 &17:20:24.18 & 57:56:26.0 & 20 & -259  & RGB star           \\
S37\tablenotemark{2}   &   &17:19:05.58 & 57:53:58.4 & 13 & -365  & RGB star           \\
M        &                 &17:20:07.44 & 57:54:32.8 & 10 & -208  & RGB star           \\
\enddata
\tablenotetext{1}{Identifications from Cudworth et~al. (2002).}
\tablenotetext{2}{Identification from Stetson (2002).}
\end{deluxetable}

\begin{deluxetable}{lcccccl}
\tablecolumns{7}
\tablenum{2}
\tablewidth{0pc}
\tablecaption{Observed Properties of Carbon Stars in Ursa Minor and Draco}
\tablehead{
\colhead{Name} &
\colhead{${\alpha}({\rm J}2000)$} &
\colhead{${\delta}({\rm J}2000)$} &
\colhead{$V$} &
\colhead{$B-V$} &
\colhead{$\langle v_r\rangle$} &
\colhead{Other} \\
\colhead{} &
\colhead{} &
\colhead{} &
\colhead{(mag)} &
\colhead{(mag)} &
\colhead{(km s$^{-1}$)} &
\colhead{} 
}
\startdata
\multicolumn{7}{c}{Ursa Minor\tablenotemark{1}}\\
UM1378    & 15:07:38.60 & 67:13:56.4 & 17.50  & 1.66  & -256.1$\pm$3.6\tablenotemark{2} & vA335, 34227 \\
UM1859    & 15:08:08.99 & 67:09:21.4 & 17.92  & 1.18  & -249.9$\pm$5.9\tablenotemark{2} & 35869 \\
UM1545    & 15:08:35.67 & 67:03:41.1 & 17.26  & 1.38  & -237.1$\pm$0.9\tablenotemark{2} & 37759 , N42 \\
UM1750    & 15:08:55.36 & 67:15:16.1 & 16.98  & 1.35  & -253.7$\pm$2.4\tablenotemark{2} & K, 33839, COS215 \\
UM1167    & 15:09:31.92 & 67:19:03.6 & 18.05  & 1.34  & -245.7$\pm$2.3\tablenotemark{2} & 32613, COS122, 70Q1 \\
UM536     & 15:11:36.35 & 67:18:07.2 & 17.58  & 1.39  & -247.6$\pm$1.4\tablenotemark{2} & 32961, JI12 \\
UM152     & 15:11:55.39 & 67:25:09.3 & 17.62  & 1.33  & -248.5$\pm$4.6\tablenotemark{2} & 30614 \\
\multicolumn{7}{c}{Draco}\\
461\tablenotemark{3}      & 17:19:42.40 & 57:58:37.8 & 17.19  & 1.74  & -299.9$\pm$1.5\tablenotemark{2} &                 \\
68        & 17:19:57.29 & 57:55:04.7 & 18.35  & 1.13  & -246$\pm$15                     &                 \\
3203      & 17:19:57.66 & 57:50:05.8 & 17.31  & 1.44  & -297.1$\pm$1.1\tablenotemark{2} & Draco C1, 22025 \\
J\tablenotemark{4}         & 17:20:00.68 & 57:53:46.7 & 17.36  & 1.59  & -300.9$\pm$0.6\tablenotemark{2} & Draco C2, 20733 \\
3237      & 17:20:33.56 & 57:50:19.7 & 17.50  & 1.55  & -280.5$\pm$0.7\tablenotemark{2} & Draco C3, 21892 \\
578       & 17:20:38.86 & 57:59:34.7 & 18.25  & 1.49  & -291.0$\pm$1.4\tablenotemark{2} & Draco C4, 18402 \\
\enddata
\tablenotetext{1}{Identifications from Cudworth et~al. (2002).}
\tablenotetext{2}{Radial velocity from Armandroff et~al. (1995).}
\tablenotetext{3}{Possible photometric variable.}
\tablenotetext{4}{Definite photometric variable.}
\end{deluxetable}

 
 
 
%
%
%
 

\begin{thebibliography}{}

\bibitem[aaron1982]{aaron1982} Aaronson, M., Liebert, J. \& Stocke, J. 1982, \apj, 254, 507
\bibitem[aaron1983]{aaron1983} Aaronson, M., Hodge, P. W., \& Olszewski, E. W. 1983, \apj, 267, 271
\bibitem[aaron1985]{aaron1985} Aaronson, M., \& Mould, J. 1985, \apj, 290, 191
\bibitem[arma1995]{arma1995} Armandroff, T.E., Olszewski, E.W., \& Pryor, C. 1995, \aj, 110, 2131
\bibitem[azzopardi1986]{azzopardi1986} Azzopardi, M., Lequeux, J., \& Westerlund, B. E. 1986, \aap, 161, 232
\bibitem[baade1961]{baade1961} Baade, W., \& Swope, H.H. 1961, \aj, 66, 300
\bibitem[bergusch1992]{bergusch1992} Bergbusch, P.A., \& VandenBerg, D.A. 1992, \apjs, 81, 163
\bibitem[canterna1978]{canterna1978} Canterna R., \& Schommer R. A. 1978, \apjl, 219, 119
\bibitem[carney1986]{carney1986} Carney, B.W., \& Seitzer, P. 1986, \aj, 92, 23
\bibitem[cote1997]{cote1997} C\^ot\'e, P., Hanes, D.A., McLaughlin, D.E., Bridges, T.J., Hesser, J.E., \&
Harris, G.L.H. 1997, \apj, 476, L15
\bibitem[cudworth2002]{cudworth2002} Cudworth, K.C., Schweitzer, A.E., \& Majewski, S.R. 2002, in preparation
\bibitem[Grillmair1998]{Grillmair1998} Grillmair, C.J., et~al. 1998, \aj, 115, 144
\bibitem[harris1996]{harris1996} Harris, W.E. 1996, \aj, 112, 1487
\bibitem[hill1998]{hill1998} Hill, G.J., Nicklas, H.E., MacQueen, P.J., Tejada, C., 
Cobos Duenas, F.J., \& Mitsch, W. 1998, SPIE, 3355, 375
\bibitem[mateo1998]{mateo1998} Mateo, M. 1998, \araa, 36, 435
\bibitem[mcclure1984]{mcclure1984} McClure, R.D. 1984, \apjl, 280, 31
\bibitem[mcclure1985]{mcclure1985} McClure, R.D. 1985, \jrasc, 79, 277
\bibitem[olsz1995]{olsz1995} Olszewski, E.W., Aaronson, M., \& Hill, J.M. 1995, \aj, 110, 2120
\bibitem[olsz1996]{olsz1996} Olszewski, E.W., Pryor, C., \& Armandroff, T.E. 1996, \aj, 111, 750
\bibitem[osterbrock1996]{osterbrock1996} Osterbrock, D.E., Fulbright, J.P., Martel, A.R.,
Keane, M.J., Trager, S.C., Basri, G. 1996, \pasp, 108, 277
\bibitem[shetrone1994]{shetrone1994} Shetrone, M.D. 1994, \pasp, 106, 161
\bibitem[shetrone2001]{shetrone2001} Shetrone, M.D., C\^ot\'e, P., \& Sargent, W. 2001, \apj, 548, 592
\bibitem[stetson1980]{stetson1980} Stetson, P.B. 1980, \aj, 85, 387
\bibitem[stetson2000]{stetson2000} Stetson, P.B. 2000, \pasp, 112, 925
\bibitem[stetson2002]{stetson2002} Stetson, P.B. 2002, in preparation
\bibitem[zinn1978]{zinn1978} Zinn, R. 1978, \apj, 225, 790
\bibitem[zinn1981]{zinn1981} Zinn, R. 1981, \apj, 251, 52
\end{thebibliography}
\end{document}